\documentclass[twocolumn,showpacs,preprintnumbers,amsmath,amssymb]{revtex4}

\usepackage{graphicx}
\usepackage{dcolumn}
\usepackage{bm}
\begin{document}
\title{Spatial separation of large dynamical blue shift and harmonic generation}
\author{Mette B. Gaarde}
\email{gaarde@phys.lsu.edu}
\author{Mitsuko Murakami}
\affiliation{Department of Physics and Astronomy, Louisiana State University,
Baton Rouge, LA 70803-4001}
\author{Reinhard Kienberger}
\affiliation{ Max Planck Institute for Quantum Optics, Garching, D-85748 Germany}


\begin{abstract}

We study the temporal and spatial dynamics of the large amplitude and frequency modulation 
that can be induced in an intense, few cycle laser pulse as it propagates through a rapidly ionizing gas.
Our calculations include both single atom and macroscopic interactions between the non-linear medium and the laser field. We analyze the harmonic generation by such pulses and show that it is spatially separated from the ionization dynamics which produce a large dynamical blue shift of the laser pulse. 
This means that small changes in the initial laser focusing conditions can lead to large differences in the laser frequency modulation, even though the generated harmonic spectrum remains essentially unchanged. 

\end{abstract}


\maketitle


In their pioneering experiments reported in \cite{LargeBlueshift2001}, Hentschel {\it et al.} demonstrated the generation of single attosecond XUV pulses by 7~fs, 750~nm laser pulses interacting with a neon gas.  As a first application, these XUV pulses were used to probe the sub-cycle time dependence of the laser electric field emerging from the neon gas. A surprisingly large dynamical blue shift was observed in this way, with a maximum value of approximately 35\% of the laser frequency, a result which was not well reproduced by theory. 

Ionization is intrinsic to all highly non-linear laser matter interactions. It contributes free electrons to the interaction medium which induces a temporal and spatial variation in the refractive index during propagation. This causes both self-phase modulation and defocusing of the field \cite{Rae1992}.
The experimental results in \cite{LargeBlueshift2001} raise a number of questions: i) Can such a large frequency modulation be caused by ionization driven self-phase modulation? ii) If so, how is it consistent with the observed harmonic cutoff energy around 90 eV, which indicates a relatively moderate interaction intensity around $5\times 10^{14}$~W/cm$^2$ \cite{KrauseCutoff}?
And iii) what is the role of the ionization dynamics in creating conditions favorable for the formation of single attosecond pulses?

In this paper we answer these questions by analyzing the temporal and spatial ionization dynamics of an intense few-cycle pulse as it propagates through a neon gas cell and generates harmonics \cite{BrabecSEWA}, using parameters similar to those of the experiment in \cite{LargeBlueshift2001}.
We show that in the beginning of the gas cell a large laser frequency modulation builds up rapidly when the intensity is still high, and then slows down toward the end of the cell because the intensity is reduced by defocusing. We find that, quite generally, the harmonic generation is spatially separated from the frequency modulation of the driving field since the XUV radiation is built up predominantly in the second half of the medium. Furthermore, we show that the frequency modulation is very sensitive to small changes in the initial focusing conditions of the laser beam, whereas the intensity of the beam after propagation and the harmonic spectrum is not. Therefore, even though the XUV {\it pulse} is an ideal probe of the laser field after its interaction with the gas, the harmonic {\it spectrum} and in particular its cutoff energy are poor indicators of the magnitude of the ionization driven dynamics. Finally we show that the ionization dynamics lead to a driving pulse at the end of the medium which has been spatially and temporally reshaped in a way that facilitates the generation of single attosecond XUV bursts.


Our description of the intense-laser matter interaction includes both the response of a single atom to the laser pulse, and the collective response of the macroscopic gas to the focused laser beam. We solve the wave equation for the driving and the generated fields by space marching through the gas, using the slowly evolving wave approximation (SEWA) \cite{BrabecSEWA}. At each plane in the propagation direction $z$, we calculate the single atom dynamics by solving the time dependent Schr\"odinger equation. The non-linear atomic response is then used as the source terms in the wave equation for marching to the next plane in $z$. This approach is described in detail in \cite{MetteAPT}. 

For the propagation of the linearly polarized driving field $E_1(t)$ we include two source terms due to the ionization of the medium \cite{Rae1992}:
\begin{eqnarray}
\frac{\partial J_{p}(t)}{\partial t} & = & \frac{e^2 N_e(t)}{m_e} E_1(t) \\
\frac{\partial J_{abs}(t)}{\partial t} & = & \frac{\partial}{\partial t} 
                                                                       \frac{\gamma(t)N_e(t)I_p E_1(t)}
                                                                       {|E_1(t)|^2},
\end{eqnarray}
where $e$ and $m_e$ are the electron charge and mass, $N_e(t)$ is the electron density, $\gamma(t)$ is the ionization rate, and $I_p$ is the atomic ionization potential. All the time-dependent quantities are also functions of the cylindrical coordinates $r$ and $z$. The plasma oscillation term $J_{p}(t)$ gives rise to a spatial and temporal variation of the refractive index which causes defocusing and self-phase modulation. The absorption term $J_{abs}(t)$ describes the loss of energy from the laser field due to the ionization of the medium. This term is small for all the cases discussed in this paper. The source term for the harmonic radiation is given by the non-linear polarization field, which is proportional to the single atom time-dependent dipole moment, calculated using the strong field approximation \cite{Lewenstein}, and the density of neutral atoms.

\begin{figure}
\includegraphics[scale=1]{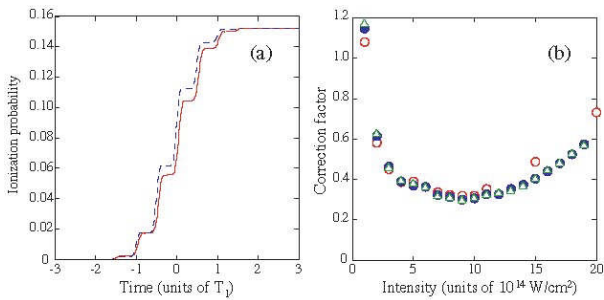}
\caption{(Color online) (a) Time dependent ionization probability of a neon atom exposed to a 750~nm, 7 fs laser pulse with a peak intensity of $10^{15}$~W/cm$^2$. $P_{vol}(t)$ is shown with solid line, $P_{ADK}$ is shown with dashed line. (b) Intensity dependent correction factor for three different pulses: 7~fs duration, cosine carrier (open circles), 5~fs duration, cosine carrier (open triangles), 5~fs duration, sine carrier (filled circles). }
\label{FigMethod}
\end{figure}

To describe the short pulse ionization dynamics correctly it is crucial to accurately calculate $N_e(t)$ and $\gamma(t)$ with sub-cycle precision. 
Our calculation of $N_e(t)$ originates in a numerical solution of the TDSE within the single active electron approximation \cite{Schafer1997}.
We define the ionization probability $P_{vol}(t)$ from the probability density of the wave function outside of a small volume around the ion core, which can be continuously evaluated during the calculation.
In Fig.~\ref{FigMethod}(a) we show $P_{vol}(t)$ (solid line) for a 750~nm, 7~fs driving pulse with a peak intensity of $10^{15}$~W/cm$^2$. Because of its short duration, the ionization probability at the end of the pulse is only about 15\%.

Ideally, we would directly couple our numerical solution of the TDSE to the solution of the wave equation and use $P_{vol}(t)$ to describe the time dependent ionization. Currently, we do not have the computational apparatus to do so. We also cannot directly use instantaneous tunnel ionization rates proposed by Ammosov, Delone, and Krainov (ADK) \cite{ADK} since these give rise to much larger ionization probabilities at high intensities than the numerical TDSE solution. However, we find that for a given peak intensity, the ionization probability $P_{ADK}(t)$ calculated from ADK rates differs from $P_{vol}(t)$ by only a constant factor $\beta$, as long as the intensity is below the (ADK) saturation intensity. We determine $\beta$ as the ratio between $P_{vol}(t)$ and $P_{ADK}(t)$ at the end of the laser pulse.
The dashed line in Fig.~\ref{FigMethod}(a) shows $\beta P_{ADK}(t)$ which is in excellent agreement with $P_{vol}(t)$.  

We next calculate $\beta(I_0)$ by comparing $P_{ADK}(t)$ and $P_{vol}(t)$ for many different peak intensities, $I_0$. This function is shown in Fig.~\ref{FigMethod}(b), open circles. 
We have calculated $\beta(I_0)$ for different pulse durations and find that it depends only weakly on the duration and the absolute phase of the driving pulse. Examples are shown in Fig.~\ref{FigMethod}(b) for driving pulse durations of 5~fs and two different carrier envelope phases. 
 Finally, we use $\beta(I_0) P_{ADK}(t)$ to calculate the source terms $N_e(t)$ and $\gamma(t)$ for each point in the non-linear medium, where $I_0$ is the peak intensity of the driving pulse at that point. The insensitivity of $\beta(I_0)$ to the duration and phase of the driving pulse ensures that this is justified even as the pulse changes shape and phase during propagation.

\begin{figure}
\includegraphics[scale=1]{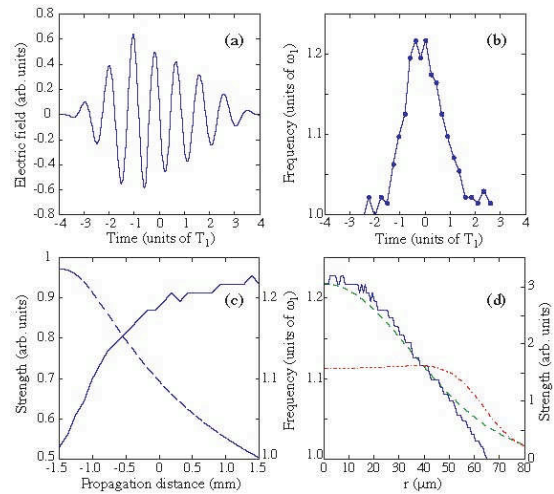}
\caption{(Color online) Spatiotemporal dynamics of frequency modulation of 750~nm, 7~fs pulse propagating through 3~mm of neon gas. See text for details.}
\label{FigFund}
\end{figure}

Fig.~\ref{FigFund} illustrates the spatial and temporal dynamics of the frequency modulation experienced by a laser pulse during propagation through a rapidly ionizing neon gas. As in \cite{LargeBlueshift2001}, the incoming laser pulse has a wavelength of 750~nm and a duration of 7~fs. The laser beam has a confocal parameter of 4.2~cm and its focus is in the center of a 3~mm long neon gas jet with a density of $5\times 10^{18}$~cm$^{-3}$. In the absence of the non-linear medium the peak intensity in the focus would be $9\times 10^{14}$~W/cm$^2$. 

In \ref{FigFund}(a) we show the on-axis electric field of the laser pulse after propagation through the neon gas. Its time dependent frequency, shown in (b),  is calculated from the separation between consecutive peaks and zeros of the electric field in (a). The pulse exhibits a large frequency chirp with a shape characteristic of self phase modulation.
The maximum blue shift which occurs near the peak of the pulse is very large, approximately 22\%, although still smaller than the 35\% shift that was reported in \cite{LargeBlueshift2001}. This will be discussed in more detail below.

Fig.~\ref{FigFund}(c) shows how this large blue shift is accumulated during propagation. The dashed line (left axis) shows the on-axis energy density and the solid line (right axis) shows the blue shift, as functions of the propagation distance $z$. At each point in $z$, the blue shift has been found as the maximum value of the instantaneous frequency as plotted in (b). At the beginning of the gas where the intensity is high, defocusing is very strong. The energy density is reduced by almost a factor of two by the end of the gas. This intensity variation in turn controls the build-up of the blue shift which accumulates rapidly in the first half of the gas and saturates towards the end. 

The solid line in Fig.~\ref{FigFund}(d) shows the radial variation of the blue shift at the end of the gas (left axis). The blue shift is largest on axis and decreases as a function of $r$. The dashed and dot-dashed lines (right axis) show the radial variation of the laser intensity at the beginning and at the end of the medium, respectively. The radial variation of the blue shift closely follows that of the incoming laser field, in agreement with the result in (c) that the blue shift is predominantly generated at the beginning of the gas.  Experimentally, the laser field is probed by the XUV pulse after refocusing both beams in to a second gas jet by a mirror placed several meters from the first jet \cite{LargeBlueshift2001}. The XUV beam in the second medium is much smaller than the laser beam and therefore only probes its on-axis frequency modulation \cite{footnote}. In principle, the radial variation of the dynamical blue shift could be probed by slightly misaligning the two beams. 

\begin{figure}
\includegraphics[scale=1]{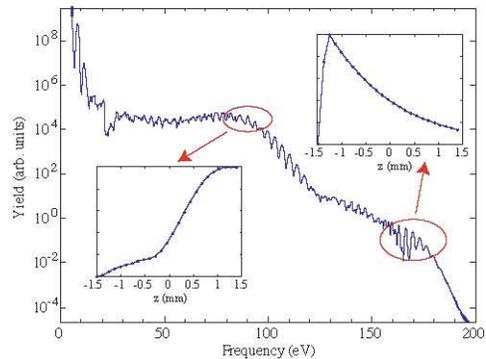}
\caption{(Color online) Radially integrated harmonic spectrum at end of the neon gas. The insets show the build-up along the propagation direction of the XUV radiation around 90~eV and 155~eV (left and right inset, respectively).}
\label{FigHarm}
\end{figure}

The ionization driven dynamics of the laser pulse has important consequences for the harmonic generation, and therefore for the formation of attosecond XUV pulses as these are synthesized from a range of harmonics near the cutoff \cite{LargeBlueshift2001}. 
Fig.~\ref{FigHarm} shows the radially integrated harmonic spectrum at the end of the neon gas. As a result of the frequency shift of the laser pulse, the harmonic structures in the spectrum are not at odd multiples of the incoming laser frequency. As a result of the defocusing of the laser beam, the harmonic spectrum exhibits two different cutoffs. The high energy cutoff around 170~eV is determined by the peak intensity of the incoming beam ($\approx 9\times 10^{14}$~W/cm$^2$) \cite{KrauseCutoff}, whereas the dominant low energy cutoff around 90~eV corresponds to the reduced peak intensity in the second half of the medium. In an experiment, it is likely that only the low energy cutoff would be observed as the high energy cutoff is orders of magnitude weaker \cite{footnoteCutoff}. 

The insets show how the XUV radiation around 90~eV (left inset) and above 155~eV (right inset) build up during propagation. The highest energies are only generated over a short distance in the beginning of the medium and are then reabsorbed through the remainder of the gas. In contrast, the 90~eV photons are generated all through the medium. The saturation of this signal at the end of the medium is due to phase matching. For longer propagation lengths one would observe a periodic increase and decrease of the yield. The total energy in a 5~eV range around 90~eV is approximately 10~pJ. 

 Fig.~\ref{FigHarm} shows that the harmonic spectrum and its cutoff energy are poor indicators of the propagation dynamics of the driving field. In a rapidly ionizing, long medium the harmonic generation is spatially separated from the defocusing and self-phase modulation experienced by the laser field. In particular, both the yield and the photon energy of the spectral cutoff are essentially decoupled from the dynamical blue shift of the laser pulse. 
 
\begin{figure}
\includegraphics[scale=1]{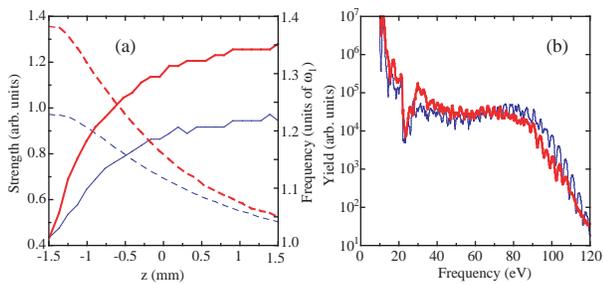}
\caption{(Color online) We compare the build-up of the frequency modulation and the harmonic spectrum generated by two laser beams with the same energy but slightly different focusing conditions. Thick lines show results of a more tightly focused beam 
than that used in Figs. 2 and 3 (thin lines).}
\label{FigDivergence}
\end{figure}

 This decoupling is further demonstrated in Fig.~\ref{FigDivergence}. We show results of a calculation where  the focusing conditions of the incoming laser beam have been slightly changed compared to Fig.~\ref{FigFund} while its energy has been kept constant. The new beam has a tighter focus with a confocal parameter of 3~cm and a peak intensity of $12.6 \times 10^{14}$~W/cm$^2$. This corresponds to decreasing the beam waist by less than 20\%. 
  
 Fig.~\ref{FigDivergence}(a) compares the propagation dynamics of the two laser beams, showing the z-dependence of the on-axis energy density and the on-axis maximum frequency shift, as in Fig.~\ref{FigFund}(c). The higher initial intensity gives rise to more free electrons, which in turn cause stronger defocusing. This means that the two beams have almost identical intensities at the end of the medium, and therefore give rise to very similar harmonic spectra, as shown in Fig.~\ref{FigDivergence}(b). However, the peak dynamical blue shift of the more tightly focused beam is much higher than before and reaches almost 35\% by the end of the medium. This answers the first two questions posed in the introduction: i) Ionization induced self phase modulation can indeed induce a 35\% frequency shift of the laser pulse in conditions very similar to those in \cite{LargeBlueshift2001}, and ii) because of defocusing, the harmonic cutoff energy is essentially decoupled from the magnitude of the blue shift. 
\\

\begin{figure}[h]
\includegraphics[scale=1]{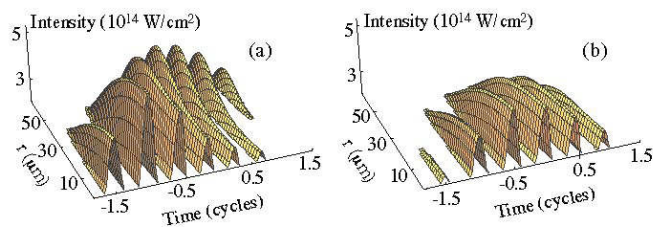}
\caption{(Color online) Spatiotemporal profile of the laser pulse at the end of the neon gas. In (a) we have used conditions as in Fig.~\ref{FigHarm}, and (b) show a laser pulse that gives rise to the same harmonic cutoff energy around 90~eV, but has a lower initial peak intensity  ($5\times 10^{14}$~W/cm$^2$).}
\label{Nearfield}
\end{figure}

To answer question iii) about the influence of the ionization dynamics on the formation of attosecond pulses, Fig.~\ref{Nearfield}(a) shows the three dimensional spatiotemporal profile of the intensity of the laser pulse at the end of the neon gas. We have used laser parameters as in Fig.~\ref{FigHarm}, and the absolute phase of the input electric field is given by a cosine driver. The strong temporal and spatial reshaping of the laser beam that takes place in the beginning of the medium results in a broad, divergent beam with a large radial phase variation, in which the peak of the pulse occurs at different times for different radial positions. In particular, this means that the pulse on axis has been shortened to a duration of less than two optical cycles (5~fs) compared to the initial 7~fs duration. 

The XUV radiation near the cutoff energy, which is used to synthesize attosecond pulses, is only generated at the highest intensities. The laser pulse shown in Fig.~\ref{Nearfield}(a) will give rise to one XUV burst on axis \cite{deBohan}, and one or several bursts off axis half a cycle later. In the nearfield, the XUV time profile will therefore in general consist of several attosecond bursts. However, the off-axis XUV radiation is generated by a strongly divergent wavefront and is much less collimated than the on-axis XUV radiation. A spatial filter placed in the farfield can therefore also act as a temporal filter, selecting only the single, collimated attosecond XUV burst generated on or close to the axis \cite{LargeBlueshift2001,footnote2}. 

To demonstrate that the ionization dynamics play a crucial role in producing the laser pulse shown in Fig.~\ref{Nearfield}(a), we show in Fig.~\ref{Nearfield}(b) the nearfield profile of a laser pulse with a lower initial intensity ($5\times 10^{14}$~W/cm$^2$ and a confocal parameter of 6~cm). This pulse gives rise to the same spectral cutoff around 90~eV. The absence of spatiotemporal reshaping of this pulse makes it much less ideal for generation of single attosecond bursts: there is no shortening of the pulse on axis, and XUV bursts generated in consecutive half-cycles of the field will have similar divergence properties because of the smooth radial structure of the beam.

In summary, we have analyzed the propagation and harmonic generation dynamics of an intense few-cycle laser pulse. We showed that the frequency modulation induced by rapid ionization in the non-linear medium can be very large, and can indeed reach a value of 35\% in conditions very similar to those in \cite{LargeBlueshift2001}. We discussed that whereas the frequency modulation is very sensitive to the laser focusing conditions, the harmonic spectrum is less so. Most importantly, we showed that the ionization driven spatiotemporal reshaping of the laser beam is very important in creating conditions favorable for the generation of isolated bursts of attosecond XUV pulses.

The authors are grateful to F. Krausz for discussions of the experiment in \cite{LargeBlueshift2001}. 
This material is based upon work supported by the Louisiana Board of Regents through grant number LEQSF(2004-07)-RD-A-09 and by the National Science Foundation through grant number PHY-0449235.


\begin{references}
\bibitem{footnote} The farfield mirror actually blocks a few mm of the central part of the laser beam. The calculated frequency modulation of the near-axis part of the laser beam after reflection on such a mirror is almost indistinguishable from the on-axis behavior as shown in Fig.~\ref{FigFund}(b).
\bibitem{footnote2} This will be investigated in detail in a forthcoming paper.
\end{references}
\end{document}